\PassOptionsToPackage{bookmarks=false}{hyperref}

\documentclass[sigconf]{acmart}

\usepackage[utf8]{inputenc}
\usepackage[english]{babel}
\usepackage{graphicx}
\usepackage{listings}
\usepackage{multirow}
\usepackage{array}
\usepackage{soul}

\begin{document}

\title{A Study of Potential Code Borrowing and License Violations in Java Projects on GitHub}

\author{Yaroslav Golubev}
\affiliation{JetBrains Research\\
ITMO University}
\email{golubev@itmo.ru}

\author{Maria Eliseeva}
\affiliation{Higher School of Economics}
\email{eliseevamary17@gmail.com}

\author{Nikita Povarov}
\affiliation{JetBrains}
\email{nikita.povarov@jetbrains.com}

\author{Timofey Bryksin}
\affiliation{JetBrains Research\\
Saint Petersburg State University}
\email{t.bryksin@spbu.ru}

\begin{abstract}
With an ever-increasing amount of open-source software, the popularity of services like GitHub that facilitate code reuse, and common misconceptions about the licensing of open-source software, the problem of license violations in the code is getting more and more prominent. In this study, we compile an extensive corpus of popular Java projects from GitHub, search it for code clones, and perform an original analysis of possible code borrowing and license violations on the level of code fragments. We chose Java as a language because of its popularity in industry, where the plagiarism problem is especially relevant because of possible legal action. We analyze and discuss distribution of 94 different discovered and manually evaluated licenses in files and projects, differences in the licensing of files, distribution of potential code borrowing between licenses, various types of possible license violations, most violated licenses, etc. Studying possible license violations in specific blocks of code, we have discovered that 29.6\% of them might be involved in potential code borrowing and 9.4\% of them could potentially violate original licenses.

\end{abstract}

\maketitle

\section{Introduction}

Previous studies have shown that a large percent of modern software consists of clones~\cite{baker1995finding, roy2008empirical, lopes2017dejavu}. In general, code clones are considered to have a negative impact on code, because they make it harder to maintain and make it harder to fix the discovered bugs~\cite{bruntink2005use, chatterji2011measuring, chatterji2013effects, mondal2019investigating}. When clones appear in different projects, they also sometimes might constitute illegal reuse of code. In open-source software, it mainly takes form of \textit{license violations}. There are two main reasons behind license violations in open-source software. Firstly, as von Krogh et al. conclude in their research~\cite{haefliger2006knowledge}, developers of open-source software reuse code in order to not focus on certain trivial tasks and because they are often limited in resources and time. Secondly, there are a lot of misconceptions about the nature of open-source software licensing, and the problem is only exacerbated by a large number of existing open-source licenses, which leads to developers not fully understanding the terms of licenses and differences between them~\cite{almeida2017software}.

In this paper, we study the distribution of licenses in Java projects on GitHub and use the time of the last modification of similar individual blocks of code to estimate possible small-scale code borrowing and license violations. Our dataset is based on the Public Git Archive~\cite{PGA} and consists of 23,378 projects with at least 50 stars and at least one line of Java code. We chose Java as the target language because of its wide spread and its wide presence in industry, where possible plagiarism is especially important to research. We carried out code clone detection with SourcererCC~\cite{SourcererCC} on the block level to capture the nature of possible small-scale copying of parts of files. The parameters of clone detection were chosen to filter out trivial pieces of code and also detect different groups of code clones.

For all the detected blocks that could constitute possible borrowing or violation, their licenses and times of the last modification were determined. Licensing information was compiled for files and for projects in the absence of the file's license. After manually inspecting each license, we discovered 94 different licenses and drew up statistics about their distribution and the relationship between the files' and the projects' licenses. 

After that, we compared the code fragments with their clones in terms of the time of the last modification, and the licenses were compared with all the older pieces of code, which allowed us to analyze the most prevalent license pairs, most probable violated licenses, and whether specific blocks of code could have come from a restrictive license. 

Our contributions are:

\begin{enumerate}
    \item Based on the Public Git Archive, we have compiled a dataset of GitHub projects with at least 50 stars that have at least one line in Java. We have performed clone detection in this dataset on the block level and compiled a list of blocks that could constitute potential license violations. For these blocks we have determined their license and their last time of modification.
    \item We have conducted the analysis of 94 licenses discovered in the dataset, including their distribution among files, and estimated possible borrowing and violations between them.
    \item We have carried out an original analysis of possible origins of specific blocks of code by studying their clones and determined that 9.4\% of blocks could constitute license violations.
\end{enumerate}

The remainder of the paper is organized as follows: in Section~\ref{related} we briefly overview existing work on large-scale clone detection in code and licensing violations, in Section~\ref{methodology} we describe the gathering of data, searching it for clones, and our analysis in greater detail, in Section~\ref{results} we present and discuss our findings, in Section~\ref{threats} we comment on the possible threats to the validity of our study, and in Section~\ref{conclusion} we draw our conclusions and reflect on future research plans.

\begin{figure*}
  \centering
  \includegraphics[width=\textwidth]{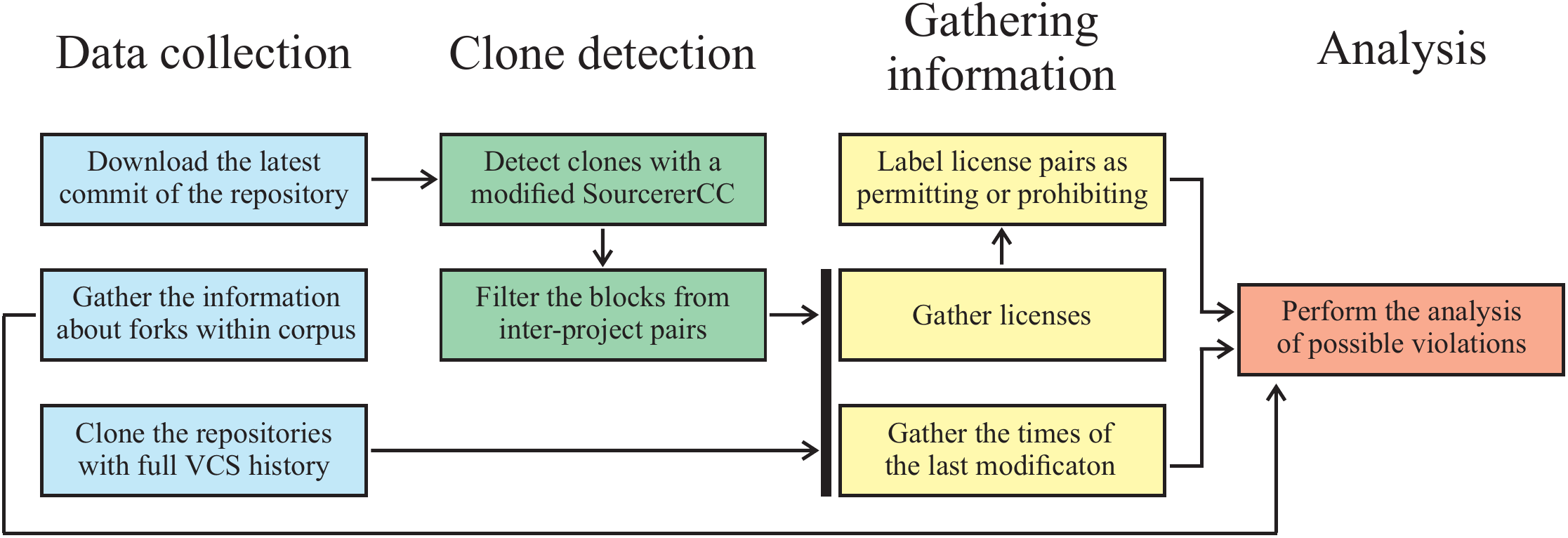}
  \caption{The pipeline of the study.}
  \label{pipeline}
\end{figure*}

\section{Background}\label{related}
\subsection{Code cloning}
A lot of previous work has focused on detecting similar code between different projects.

Some earlier works~\cite{mockus2007large, mockus2009amassing} use a very straightforward approach for detecting possible copies of code in large systems: they study names of directories and files and identify directories that share several source files (the amount of the shared files acts as a threshold). While this approach can correctly identify exact copies of projects and scales well for very large datasets, such methods are inapplicable when it comes to determining actual plagiarism and licensing violations in open-source software, because similarities need to be detected in the code itself.

A more thorough way of detecting similar code is to look for similar files and search for file-level similarities. Ossher et al.~\cite{ossher2011file} study over 13,000 Java projects and use three various techniques to discover three degrees of file-level cloning in their dataset: firstly, they compare MD5 hashes of files to determine the exact copies, then they discover files that have the same Java fully qualified names, and, finally, to determine similar code with more alterations, the authors also compare name-based fingerprints of files. The authors have determined that in the excess of 10\% of files are clones. This method is more delicate, since it takes into account the content of files, however, with a rapid development of various clone detection tools~\cite{rattan2013software, ain2019recent}, one would like to also implement them in such a study.

Clone detection was employed by Lopes et al.~\cite{lopes2017dejavu} on an enormous dataset of several million projects in four languages: Java, C++, Python, and JavaScript. The authors detect clones on several levels: firstly, they also compare MD5 hashes of files to discover exact matches, then they compare the hashes of tokens in files to discover permutations in files, and, finally, they use SourcererCC~\cite{SourcererCC} clone detection tool to discover near-miss clones on a file level. The authors present a wide range of statistics and in the end discover that of 428 million files in their corpus, only 85 million are unique, showcasing the prevalence of code cloning.

However, code cloning can occur on a scale smaller than files. Roy and Cordy use an improved version of the NICAD~\cite{roy2008nicad} clone detection tool to conduct a study of \textit{function-level} code clones between systems~\cite{roy2010near}. The authors discover a large number of exact clones between systems and an even larger amount of near-miss code clones. 

Taking another look at the problem of code cloning and its possible flow, Yang et al.~\cite{yang2017stack} compare a large dataset of Python GitHub projects to code snippets on StackOverflow on a block level and discover certain cases of code from StackOverflow making its way onto GitHub. In this study, the authors also used SourcererCC for its ability to process blocks of code on a method level.

Overall, the problem of clone detection on a level of code segments requires further research and is especially interesting because a lot of code cloning (and, consequently, possible license violations) occurs with pieces of code and not entire files.

\subsection{Licensing violations}

The issue of discovering possible illegal reuse of code requires not only detecting clones, but also correctly identifying licenses in code and analyzing various cases of licensing violations.

As it was mentioned before, there are a lot of open-source licenses. For a quantitative study, one might turn to Vendome~\cite{vendome2015large}, who analyzed 16,221 Java projects and discovered at least 25 different open-source licenses within them using a tool called Ninka~\cite{Ninka}. In that work, the trends of licenses popularity are also demonstrated, showing, for example, the rise of popularity of Apache-2.0 license in the years 2002--2012. The dynamics of licensing is closely studied by Vendome et al.~\cite{vendome2015and}, who survey developers as to when and why they adopt and change licenses in the code; the reasons include guidelines typical to their communities, purpose of usage, and the usage of third-party libraries. 

Another very popular open-source license is GNU General Public License (GPL), however, it is much more restrictive than Apache-2.0. Research was conducted~\cite{german2010understanding} that discovered license incompatibility issues with it. Code licensed under GPL is often incorrectly reused in more permissive licenses.

In another research~\cite{german2009code}, the possible types of licenses are also discussed and a quantitative analysis of licensing in Linux, FreeBSD, and OpenBSD is conducted. Here, authors note the possible cases of multi-licensing that occurs during the migration of code.

A very specific empirical study is carried out by Mathur et al.~\cite{mathur2012empirical}, where authors analyze 1,423 projects from Google Code, search them for clone files, draw up statistics of licenses usage in the corpus and then present a case-by-case descriptions of detected licensing violations.

Wu et al.~\cite{wu2015method, wu2017analysis} analyze licensing inconsistencies in Debian~7.5 by implementing file-level clone detection and using Ninka to determine their licenses. The authors point out cases when identical files exist in two packages that differ only by the license in the header, which constitutes licensing inconsistency. The authors categorize inconsistencies as license additions, removals, change, and also upgrades and downgrades in the case of GPL licenses.

Zhang et al.~\cite{zhang2010automatic} propose a tool called \textit{LChecker} that gathers the information about the license of each file in a given project, tokenizes it, and then looks for similar files via Google Code Search. The licenses of the discovered files are compared to the license of the original file by a voting mechanism, and possible license inconsistencies are returned. The authors used this tool in several experiments and discovered several new license violations.

In general, one might see that licensing is a complex and non-trivial subject that requires a lot of delicacy. A lot of tools have been created for the automatic determining of the file's license and used to discover license incompatibilities, however, this field also still requires further research.

A very peculiar study that combines function level clone detection between systems and licensing violations is carried out by Romansky et al.~\cite{romansky2018sourcerer}, who study possible licensing violations between Python modules and StackOverflow posts. The authors use SourcererCC to detect clones on the block level and a modified version of Ninka to detect licenses in files. The study concludes that code migration co-exists with constant relicensing of code that often violates the original licenses and that StackOverflow posts often contain code from licenses that are incompatible with StackOverflow's license. 

There is an interest in conducting a similar large-scale analysis of code fragments in a large corpus of code to compile the list of all present open-source licenses and their distribution as well as to estimate the possible degree of code borrowing and licensing violations in the code by analyzing the amount of clones between these licenses. In this paper, we aim to carry out such a study using Java as our target language and GitHub as the platform. Let us now walk through each step of the pipeline, presented in Figure~\ref{pipeline}.

\section{Methodology} \label{methodology}

\subsection{Data collection}

Our corpus was based on the Public Git Archive (PGA)~\cite{PGA}. PGA is a large dataset that was composed in the early 2018 using GHTorrent~\cite{GHTorrent} metadata and consists of all GitHub projects with 50 or more stars. PGA allows to filter repositories by language, which we used to extract the list of all the projects with at least one line written in Java, 24,810 projects overall.

While PGA supports a convenient and easy to implement mechanism for a bulk download of projects, the analysis of possible borrowings and violations requires an access to full and up-to-date Version Control System (VCS) data, namely, the timestamps of the most recent modifications to certain lines of code. Therefore, each repository was downloaded with full history, as well as downloaded as a zip-file of the most recent commit for processing with a clone detection tool. Due to some projects being deleted or made private, 286 projects were no longer available for download. Also, for every project, their current full name was requested via GitHub API, and it turned out that due to relocations, certain projects names that are different in PGA now lead to the same repository. In such cases, a single repository was left out of the repeated ones, which excluded 807 more projects. Finally, 339 repositories did not contain any lines in Java anymore at the time of the download, so they were not present in the results as well. Therefore, the final dataset consisted of 23,378 Java repositories with their histories up to June 1st, 2019, the full list is available.\footnote{Dataset: \url{https://doi.org/10.5281/zenodo.3608211}} Collectively, both their full history and their zips took up 1.55 TB of hard drive space. 

In order to consider forks in our analysis (since clones between forks most likely do not constitute code borrowing), GitHub API was also used to gather a list of all forks for all the projects in our dataset, after which it was determined which projects within the dataset are connected.

PGA also stores the information about the project's license. The information was gathered using Go License Detector\footnote{Go License Detector: \url{https://github.com/src-d/go-license-detector}}, a tool that scans the directory for license files and outputs the probabilities of it being under a certain license. The tool supports the entire SPDX license database\footnote{SPDX license list: \url{https://spdx.org/licenses/}} and for any particular project can list one possibility, several possibilities, or nothing if the project has no license file at all. In our work, we consider the project's license to be the one that has the highest probability according to PGA, and if the project has no license, it is listed as \textit{GitHub}, which means that \textit{All Rights are Reserved} (that includes a large number of projects, almost 30\% of the full PGA).

The list of the projects was composed and passed on to a modified version of SourcererCC.

\subsection{Clone detection}

SourcererCC~\cite{SourcererCC} is a clone-detecting tool that implements a complex mechanism for comparing blocks of code that uses reversed indexing and various heuristics to detect different types of clones, while also having a reasonable operating time. SourcererCC is token-based, defining a token as a programming language keyword, a literal, or an identifier. The tool parses the files and tokenizes the data, and then uses this tokenization to compare pieces of code to detect possible clones. In our research, we have chosen this particular tool for two reasons. 

Firstly, it supports detecting clones on the method level, which suits our interest of detecting subtle, small-scale copying of code, which stems from the fact that users rarely copy entire files, they often copy only parts of it. As a unit of code, a method provides a balanced approach to clone detection: it preserves the semantics of the code, while also not being too general and avoiding missing the details.

Secondly, we have managed to improve the tool by upgrading its two main stages:

\begin{itemize}
    \item The tokenization was rewritten to Python3 and modified to open each file with UTF-8 encoding, solving the problem of the tool omitting files with non-ASCII characters in them, which had a significant impact on our study, because a large corpus inevitably has a lot of files with commentaries in various languages, for example, we faced files with a lot of Chinese characters in them.
    \item Clone detection was performed in several parallel instances with various parameters (Similarity Threshold and Lower Token Length Threshold). The main search was performed with 75\% Similarity Threshold and the Lower Token Length Threshold of 19 tokens to filter out trivial pieces of code, these parameters were determined empirically in preliminary experiments. In their research, Saini et al.~\cite{saini2018cloned} use similar values of 70\% and 25 tokens. We have also used other parameter pairs to discover larger blocks that could be considered clones due to possible sub-block correlations. The process requires a significant amount of computational resources, but is well-suited for large-scale studies like this one that search for clones only once and value fullness over cost-effectiveness. The parameters of the search were chosen so that the precision of the results was 90\%, which was determined by manually labeling a statistically significant sample of clone pairs.
\end{itemize}

The results of clone detection were merged together, the output data consisted of a list of all clone pairs and the information about every block, including the project it came from, the file address in the project, and lines in the file. 

A separate clone-finding process was also carried out with the Similarity Threshold of 100\% to find exact copies of code. 

\subsection{Gathering information}\label{statistics}

\subsubsection{Licenses and Blames of files}\label{blames}
First of all, a set of all blocks (and files they belong to) that appear in \textit{inter-project} pairs was created from the detected clone pairs. Studying license violations requires us to collect two types of data for these fragments: licenses and the time of the last modification, which allows us to presume what code could have been copied from where.

While PGA contains information about the project's license, this doesn't provide a desirable level of accuracy. A lot of large projects consist of several parts, often developed by different teams and at different points in time. Even more importantly, they could incorporate various libraries and tools within a project, which leads to different files having different licenses within a single project. Sometimes, a single file even has several licenses that a user might choose from. Even though such cases are naturally rather rare, it is of interest to study them in greater detail.

In order to determine the licenses in files, we used Ninka~\cite{Ninka}. Ninka is a tool written in Perl that uses a sentence-based approach to parse the top part of each file and match it to the known licenses. Similarly to Go License Detector, the output can contain a single license, several licenses, ``Unknown'', or ``None''. ``Unknown'' relates to the case when some words and sentences in the header are considered relevant, but cannot be matched to an exact license. That is often the case when a file has a simple copyright line in the header, which often repeats the project's license or when the existing license is formulated in an unusual way. Overall, for any given file, we pick the license that Ninka considers to be the most probable, and pick the project's license if Ninka outputs ``Unknown'' or ``None''. There might be more complex relations between the licenses of the file and the project, but in this research, we assume that if the license of the file is different from the license of the project, that has a certain reason for it that must be accounted for.

Since Ninka outputs licenses differently than PGA, we have compiled a list of all the existing outcomes that appear in our files (both PGA and Ninka) and manually checked every one of them. For every entry, we have reviewed from 3 to 20 files. For the licenses that are the same, we have unified the styles of output, and we have also fixed discovered inaccuracies. In the end we have compiled a list of licenses by number of files with them and a list of files by the amount of licenses per file, which are presented in Section~\ref{results:licenses}.

The second step in analyzing possible code borrowing is determining the timestamp of the last modification to the code. For this, we make use of the full VCS history we downloaded. Git allows to gather the necessary information by using the \textit{git blame} command, the output of which includes each line of the file, as well as its last time of modification (date and time) and the author of this modification. This system is well-suited for our task of suggesting possible violations on the block level, since the information is not generalized for the entire file.

An extensive research of using \textit{blame} commands to distinguish between the originals and the clones in source code was conducted by Krinke et al.~\cite{krinke2010distinguishing, krinke2010cloning}

\subsubsection{Transforming the data}\label{sec:blocks}
The information about the code blocks was obtained in the following way. A block's license was considered the same as the containing file's license (or the project's license in the case of its absence). As for the timestamps, we used the following algorithm. Since SourcererCC stores a line range in the file for every single block it tokenizes, a list of timestamps of every line of a given block was extracted. After that, the mode (the most frequent date) was calculated for that list and was considered to be the block's overall timestamp. If the list has several equal modes, we chose the most recent one.

The reasoning behind this algorithm is the following. Imagine that a clone pair of blocks was found of 25 lines each. One of them has the same time of last modification for every line, somewhere in 2017. The second one has 22 lines from 2015 and 3 lines from 2019. Even though the block as a whole was updated more recently in the second case (and on a file level, Git would show 2019 as the time of the last modification), statistically it makes more sense to consider 2015 as the time of last modifications, since it is more probable that the latest change didn't affect the possible detection of the clone. There are a lot of different scenarios for changing just several lines of code into a method, like renaming a variable or fixing a small bug. And borderline cases, where exactly half of the block was written at one point and half was written at another, turned out to be rather rare in our dataset.

Despite the fact that \textit{git blame} provides the timestamp down to a second, the level of temporal granularity was chosen to be a day as a precaution against detecting possible code from the same contributor in different projects.

\subsection{The analysis}\label{sec:analysis}
\subsubsection{Possible code borrowing and license violations}\label{sec:analysis_borrowing}

Several algorithms for detecting license violations on the block level were considered. All of them revolve around viewing the results of the clone detection as a graph structure, with blocks being the vertices and the clone pairs being the edges. Such a graph has a fairly complex structure, with a large amount of small connected components (down to a single pair) and long strings of connected vertices. 

\begin{figure*}
  \centering
  \includegraphics[width=\textwidth]{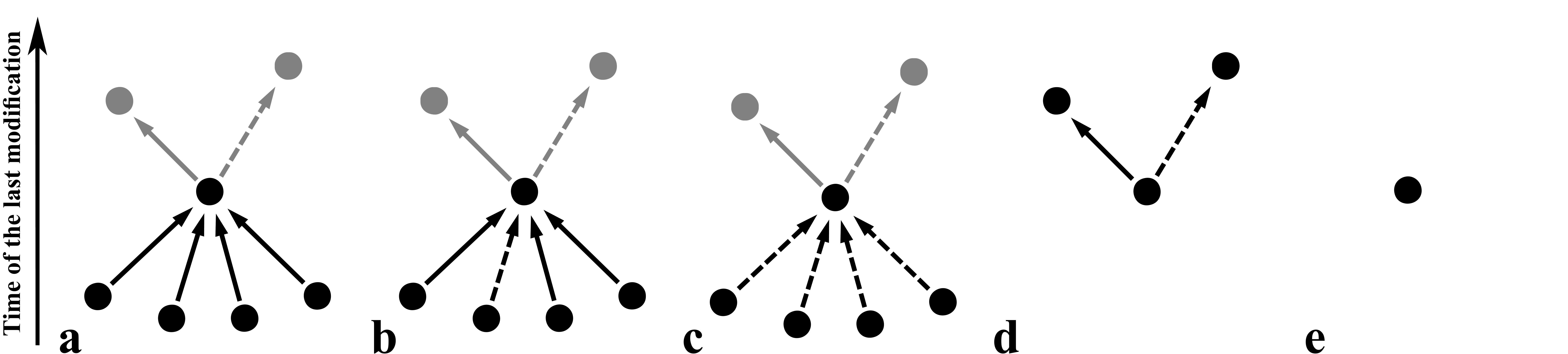}
  \caption{Five types of blocks during the analysis of the neighbors: \textbf{a)} Strong violation; \textbf{b)} Weak violation; \textbf{c)} Legal borrowing; \textbf{d)} Origin block; \textbf{e)} Unique block. The vertical axis indicates the time of the last modification of the blocks, straight lines represent prohibited copying, dashed lines --- permitted copying, grey lines indicate that connections are optional for this type.}
  \label{types}
\end{figure*}

Three types of connections between blocks in the graph have been terminated:

\begin{itemize}
    \item Firstly, we do not take into account pairs between forks. It should be noted that not only the connections between the original project and its fork were considered, but also the connections between two forks of the same project and in general between any projects that have a common parent project on a certain level. From our experience, the pairs between them are more probable to not be clones but rather the products of the same origin.
    \item By checking the resulting inter-project pairs for any anomalously large amount of pairs between specific pairs of projects, we have discovered several mirrors --- projects that are not forks, but were copied at some point, have the same contributors and, therefore, also do not constitute code borrowing. The largest such example is Jython.\footnote{Jython copies: \url{https://github.com/jythontools/jython} and \url{https://github.com/jython/jython3}}
    \item We also do not consider projects that belong to the same user or organisation, since we assume that almost always their rights are shared.
\end{itemize}  

We then took the graph vertices one by one and analyzed all of their neighbors. The time of the last modification for the vertex was compared to every one of its neighbors, which allowed us to turn an undirected graph into a directed graph where edges become arrows from the older vertex to the younger vertex. Connections with the same date are not considered at all, because such connections are not very frequent and might correspond to undetected mirrors of projects.

For every predecessor in the graph, a pair of licenses (predecessor \(\rightarrow\) current block) was registered to create a list of all possible license pairs of code borrowings. After this process commenced for every block, a ranged list of possible license pairs was drawn up. Naturally, some possibilities are much more common than other, so in order to consider possible violations, we have labeled the pairs as follows: we have started with the most popular license pair and worked our way down until 99\% of the potential borrowings were labeled. The license pairs were labeled as either \textit{permitted} or \textit{prohibited}. In our labeling, we assumed the best possible conditions, for example, that the copyright is always saved and present and that the original license is always present somewhere in the appendix of the main license or in the directory of the receiving file. Basically, the following cases constitute possible violations:

\begin{itemize}
    \item Copying from files with no license (GitHub license), since they require explicit permissions from the author. It might be the case that the receiving project has that permission, but since we cannot know that, we consider such cases as constituting a possible violation.
    \item Copying to files with no license (GitHub license), since virtually all open-source licenses require the incorporating project to also be open-source.
    \item Prohibited copying from copyleft licenses (like GPL) to permissive licenses.
    \item Certain unique incompatibilities between older versions of licenses.
\end{itemize}

All the license pairs after that (that constitute the remaining 1\% of the possible borrowings) were labeled as permissive, because it required too much manual work to label the vast majority of the possible cases that have almost no effect on the general picture of the possible violations in the corpus. The only exceptions are borrowings from and to files with no licenses, such license pairs were all considered prohibited.

Based on this data, various statistics was collected: the most violated licenses, the most violating licenses, the most prominent license pairs, and borrowings between the most popular licenses. 

\subsubsection{Possible origins of blocks}\label{sec:analysis_blocks}

After the labeling was complete, the analysis of the neighbors of each block was repeated, but now, for every block, the connections to its predecessors were considered as either permitted or prohibited as per the labeling.

In total, we can define five different configurations (see Figure~\ref{types}) that all blocks fall into. Their relationship towards license violations is determined by the \textit{Violation coefficient}:

\begin{equation} \label{eq:1}
P_{viol} = \frac{N_{prohibited}}{N_{all}},
\end{equation}

where \(N_{prohibited}\) is the number of prohibited predecessors of a block and \(N_{all}\) is the total number of predecessors. Two of the configurations constitute possible license violation:

\begin{itemize}
    \item \textit{Strong violation} (Figure~\ref{types}a) occurs when a block has predecessors and all of the connections are prohibited by their licenses. That means that this piece of code came from one of the restricting licenses, and \( P_{viol} = 1 \).
    \item \textit{Weak violation} (Figure~\ref{types}b) occurs when a block has predecessors and only some of the connections are prohibited, meaning that this code \textit{might have} come from a restricting license, but also could be coming from a permitting licence, \( 0 < P_{viol} < 1 \).
\end{itemize}    

Three other configurations that appear in our graph all have \( P_{viol} = 0 \):

\begin{itemize}
    \item \textit{Legal borrowing} (Figure~\ref{types}c) occurs when the code has one or several predecessors, but all of the connections are permitted.
    \item \textit{Origin block} (Figure~\ref{types}d) is a block that has no predecessors but has successors, indicating that it takes part in the cloning process, but only as a source.
    \item \textit{Unique blocks} (Figure~\ref{types}e) are blocks that do not engage in any cloning. Technically, they are not in the output of SourcererCC, however, they are present in the graph as isolated vertices. To find them, one needs to look through the tokenization data and filter all the blocks that were tokenized and that pass over the minimal Token Length Threshold, but are not present in the resulting pairs. One might also separately count the blocks that have no clones at all, anywhere, and blocks that only have clones within a project, with a forked project, or within the same author and thus cannot constitute a violation.
\end{itemize}

\section{Results and discussion}\label{results}

The research was carried out using Amazon Web Services (AWS). In total, eight servers with Intel Xeon Platinum 8175M @ 2.50 GHz and 30 GBs of RAM were used. One of the servers required 2 TB of hard drive space to store all the downloaded projects, while the rest were only used for detecting clones (which doesn't require source files) and gathering the necessary code data.

The gathering of the dataset took about a day, and the clone detection took a little over two months of continuous calculations. In total, the dataset was tokenized into 38,617,427 unique blocks of code. 11,762,703 blocks of code passed the threshold of 19 tokens, of which 7,601,738 engaged in the cloning process (64.6\%). In total, 1,163,989,420 clone pairs were detected, which came from 20,824 different projects, meaning that 2,554 projects did not have any clones larger than the threshold at all. Out of these pairs, 560,656,419 were inter-project (48.2\%).

324 projects in our dataset have forks that are also in our dataset. We have also specifically checked the dataset for the forking cases of \(A \rightarrow B \rightarrow C\), where all three projects are in our dataset. There are only three such cases in the corpus. The reason for that is that the corpus only consists of projects that had 50 stars at the moment of its creation, and forks of projects tend to have less stars than originals. So there can be a lot of \textit{forks of forks}, but very few of them will have 50 stars or more.

\subsection{Licenses}\label{results:licenses}

After filtering file names with certain special characters that would not allow Ninka to process them (which was only about a thousand files), 557,553,075 pairs continued on to the data gathering stage. 

Overall, 3,844,515 blocks of interest were extracted that originated from 897,620 files of interest. For each of these files, we collected the licensing information with Ninka and \textit{git blame} output. All blocks of interest were subsequently correlated with their last time of modification using the algorithm described in Section~\ref{sec:blocks}.

As previously mentioned in Section~\ref{blames}, the output of Ninka and PGA is different, so all the obtained licenses of the files were drawn up and checked. There were a total of 347 different variants, and all of them were checked manually. The rarer licenses were checked through especially thoroughly to make sure that all the licenses that are listed in the corpus actually appear in it at least once and are distinct from other ones.

After the processing, we have discovered 94 different licenses in the dataset, the full list is available.\footnote{The list of discovered licenses: \url{https://doi.org/10.5281/zenodo.3665281}} A lot of these licenses fall into the same family. For example, there are 8 variants of GPL license and 11 variants of BSD licences, a lot of which have very small and specific differences between them, like the prohibition of use in ``the design, construction, operation or maintenance of any nuclear facility'' (BSD-3-Clause-No-Nuclear-License).\footnote{BSD-3-Clause-No-Nuclear-License: \url{https://spdx.org/licenses/BSD-3-Clause-No-Nuclear-License.html}} The most licenses in one family belong to Creative Commons licenses, of which there are 13.

\begin{figure}
  \centering
  \includegraphics[width=2.8in]{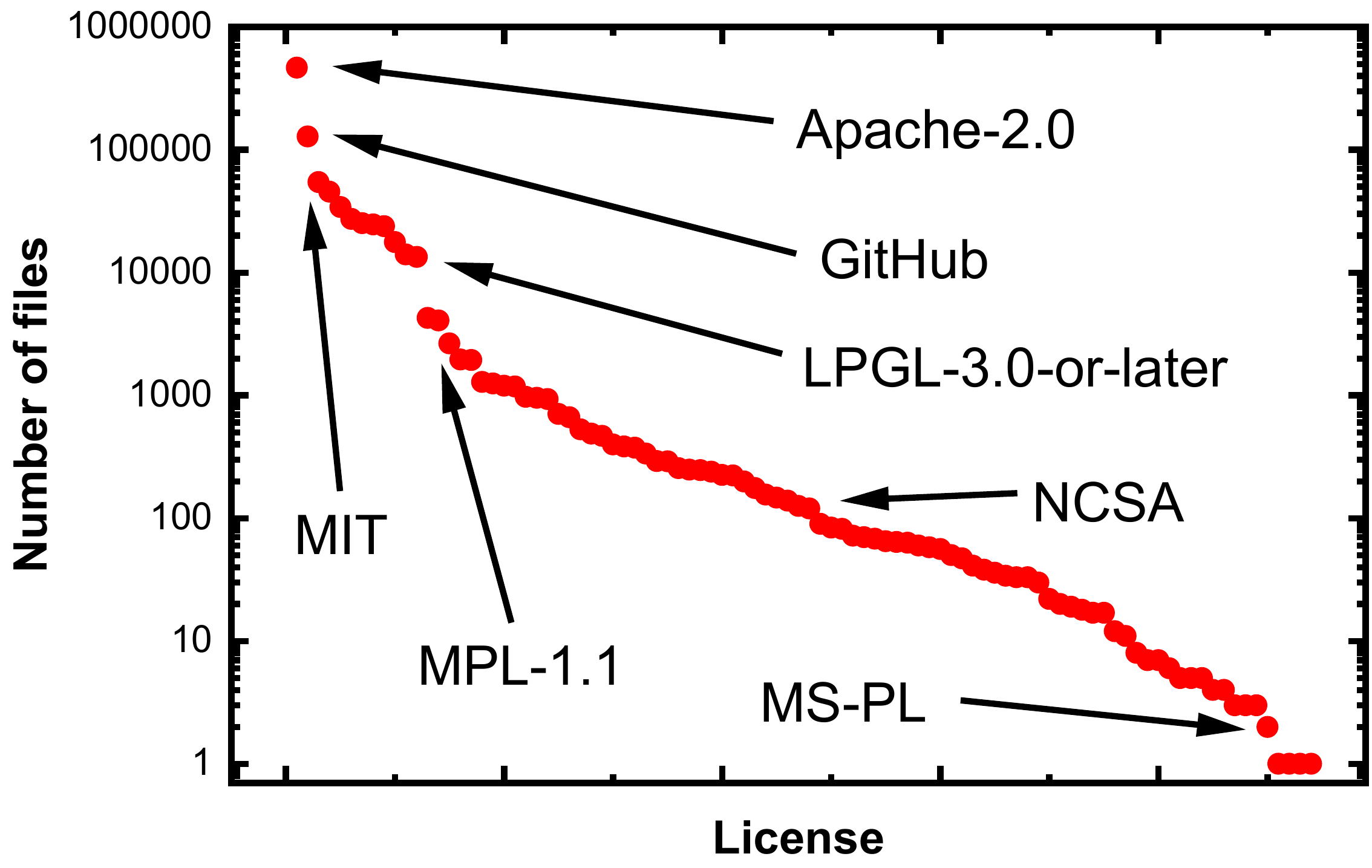}
  \caption{The number of files with various licenses.}
  \label{licenses}
\end{figure}

The distribution of files covered by various licenses is presented in Figure~\ref{licenses}. Three areas can be discovered on the graph: two entries both have more than 100,000 files, ten more licenses have between 10,000 and 100,000 files and a lot more smaller licenses have less than 10,000 files.

\begin{figure}
  \centering
  \includegraphics[width=2.8in]{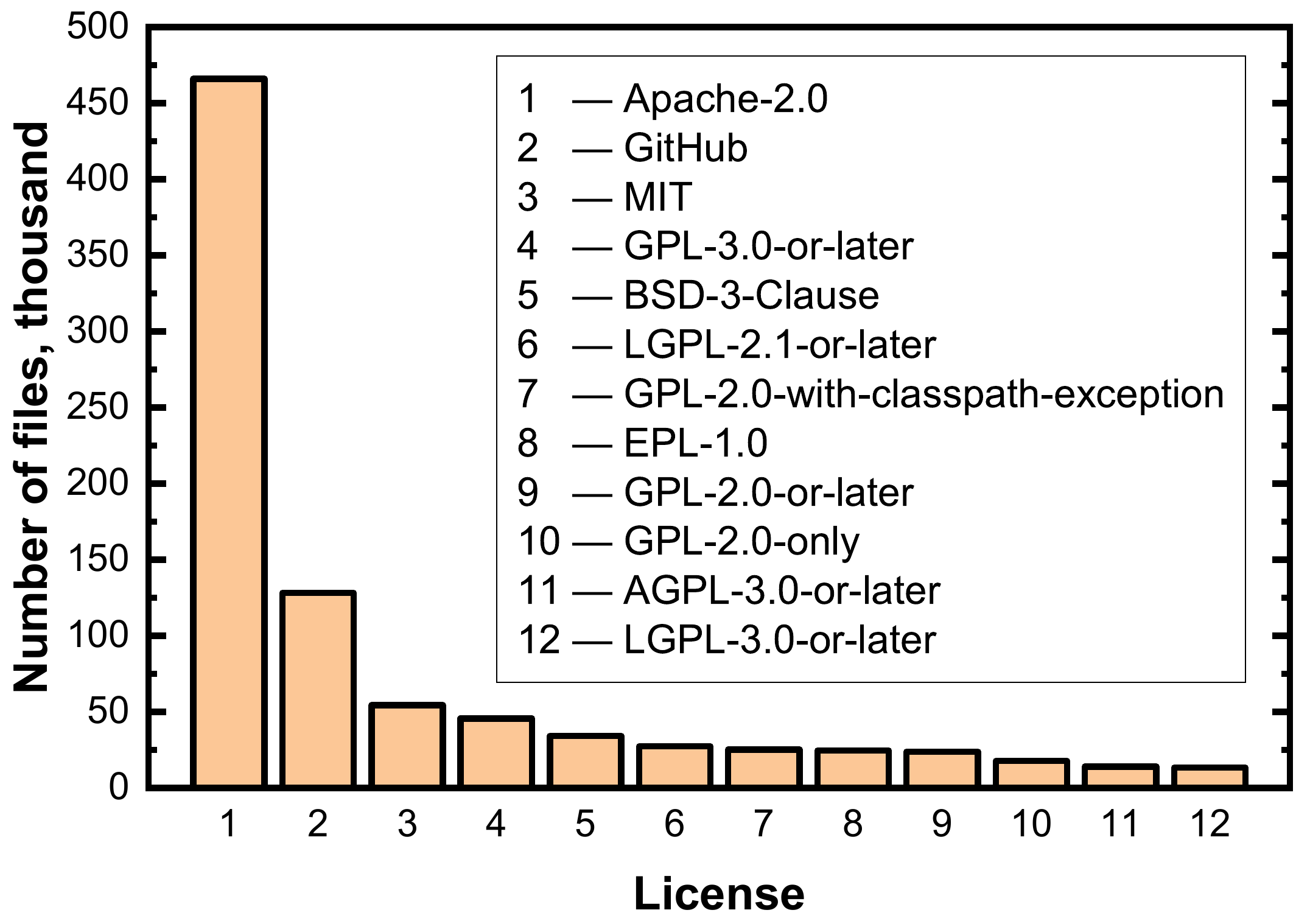}
  \caption{Twelve most popular licenses by the amount of files.}
  \label{top12}
\end{figure}

Let us take a closer look at the most popular discovered licenses. They are shown in Figure~\ref{top12}. In files that we have analyzed, the most popular license by far is Apache-2.0. It covers more than half of the files (51.9\%) and greatly surpasses all the other licenses, which is consistent with recent research~\cite{vendome2015large}. The license is very popular because of how permissive and detailed it is.

The second most popular license is actually not a license, but its absence. A significant amount of Java code on GitHub comes from files with no license (14.3\%). When a developer uploads code to GitHub and does not provide any license with it, then \textit{all rights are reserved} and the borrowing of code requires explicit permission from the author. Using GitHub as a platform implies agreeing to its Terms of Services that allows free viewing and forking of the code, however, free copying is not allowed. Therefore, it can be expected that such code will be especially prevalent in possible violations. In the following figures, such case is denoted as \textit{GitHub}.

Among other popular licenses, there are two other very popular permissive licenses: MIT (6.1\%) and BSD-3-Clause (3.8\%). Both of them are much shorter and simpler than Apache-2.0, but convey a similar idea. They allow free modification and distribution of the code, and MIT allows sublicensing. A different case are GPL licenses. A lot of licenses, including Apache and BSD, have versions, but with GPL it is especially  important. All together, different versions of GPL license cover almost 115,000 files (12.6\%) in our dataset, making them collectively the third most popular one. Unlike the above-mentioned licenses, GPL licenses have strong copyleft requirements, meaning that the modified versions of the code must be released under the same license. In the similar vein, GNU Lesser General Public Licenses (LGPL) deal with ``weak copyleft'' and GNU Affero General Public License (AGPL) is developed for server-side applications. 

\begin{figure}
  \centering
  \includegraphics[width=2.8in]{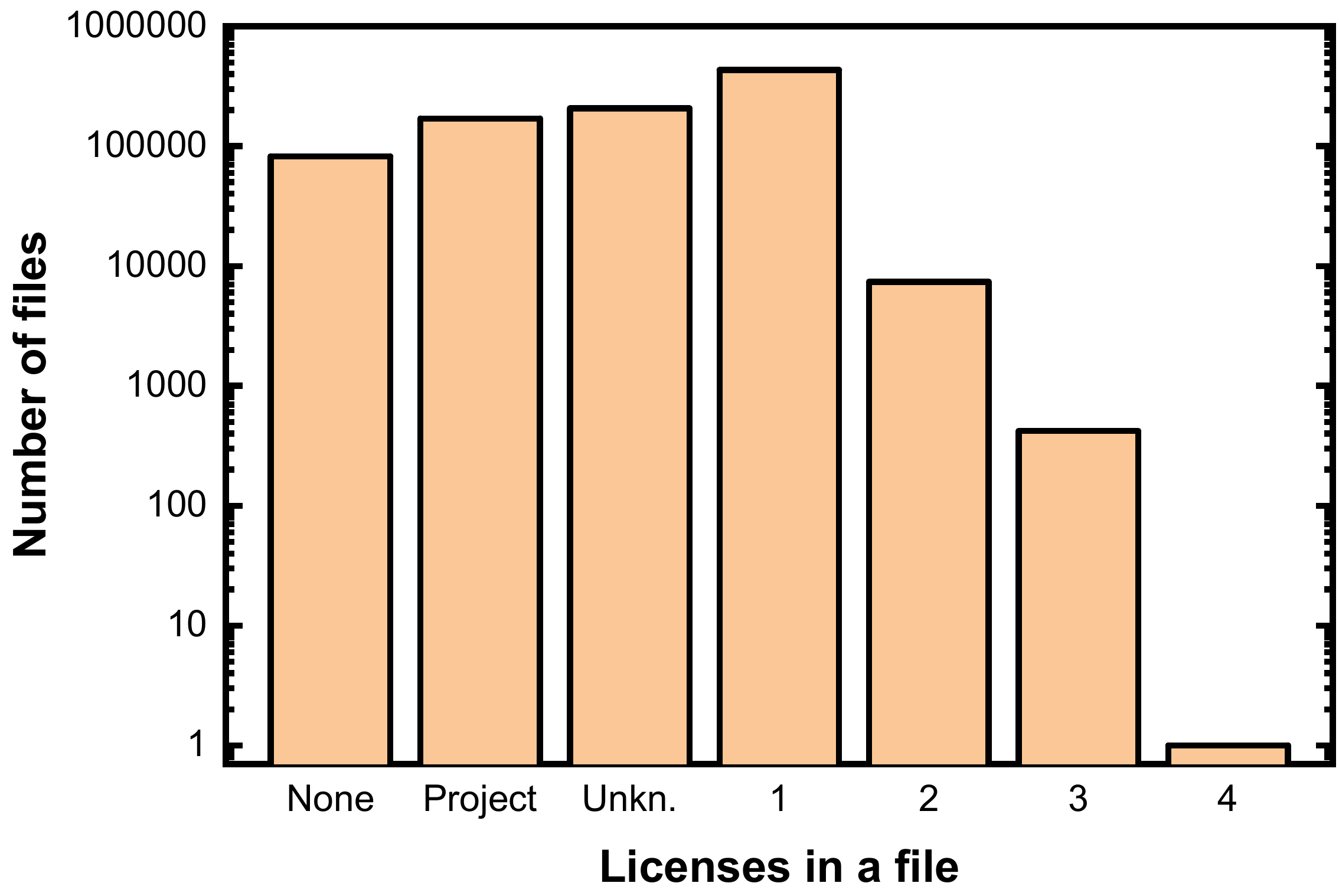}
  \caption{Files by the licenses that cover them. Numbers indicate the amount of licenses in the file's header, \textit{Unkn.} relates to the case when Ninka is unsure of the license, \textit{Project} means that the file has no license in the header but is covered by the projects' license, and \textit{None} means that neither the header nor the project have a license.}
  \label{multilicense}
\end{figure}

Another interesting case to look at is multi-licensing. In our work, we did not take into account the multi-licensing on the projects' level, because we assumed the license to be the one predicted as a most probable by PGA. Ninka, however, can not only propose several \textit{possibilities} of licenses in files, but also correctly detect several licenses. So here we talk specifically about multi-licensing within a file.

The distribution of the amount of files on the conditions of licensing is shown in Figure~\ref{multilicense}. Multi-licensed files are not frequent in the corpus (only 0.9\%). There are 74 combinations of two licenses, of which two combinations are the most frequent (an order of magnitude more present than others). The first combination is CDDL (Common Development and Distribution License) + GPL-2.0-with-classpath-exception, which is almost always copyrighted by Sun Microsystems or Oracle and is specific to them, the second combination is Apache-2.0 + GPL-2.0-or-later. There are eight different combinations of three licenses, with the largest being GPL-2.0-or-later + LGPL-2.1-or-later + MPL-1.1 (Mozilla Public License), which is always presented as a single block with all the licenses listed in the header. There is a single file in the entire corpus that mentions four licenses in the header. They consist of the above-mentioned pair CDDL + GPL-2.0-with-classpath-exception, but there is also the header of Apache-2.0, which is the license of the entire project, and a notice of CC-PDDC (Creative Commons Public Domain Dedication and Certification).

\subsection{Possible borrowings of code}

As described in Section~\ref{sec:analysis_borrowing}, in order to detect possible code borrowings, we filtered out clone pairs between forks, mirrors and the same author. Using the data from the separate 100\% similarity run we have conducted, we have discovered that among the remaining pairs, 11.7\% constitute exact clones, indicating that there is a significant amount of the same code between unrelated projects in our dataset.

For the pairs where the blocks differ in their last time of modification by at least a day, license pairs \((\text{older block} \rightarrow \text{younger block})\) were analyzed. All the possible license pairs were ranged from the most frequent down. Overall, the first 176 pairs covered 99\% of all the possible borrowings, so they were manually labeled as either permitted or prohibited. 

\begin{figure}
  \centering
  \includegraphics[width=2.8in]{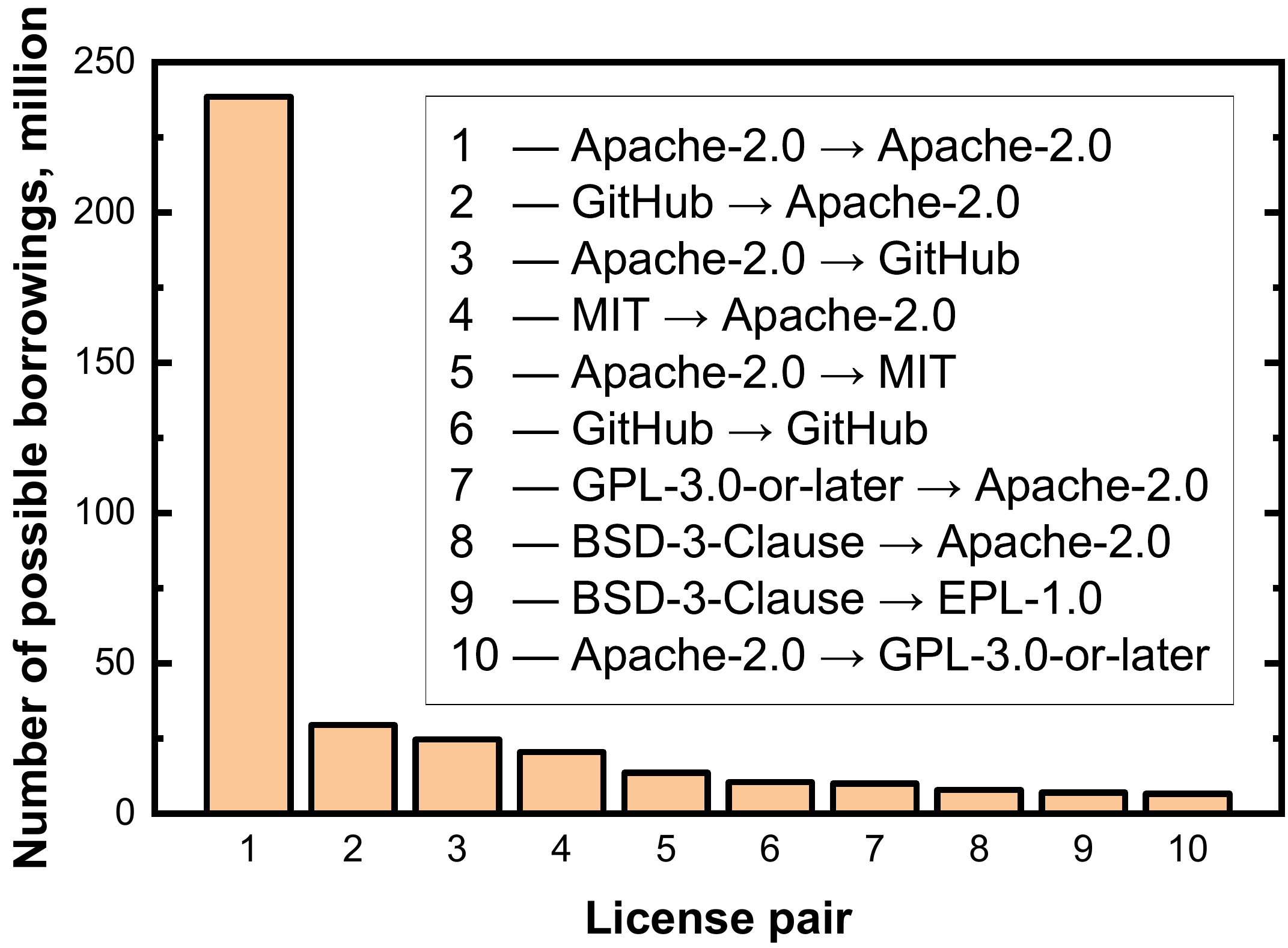}
  \caption{Ten most popular license pairs by possible borrowings.}
  \label{top10borrowings}
\end{figure}

\begin{figure}
  \centering
  \includegraphics[width=3.3in]{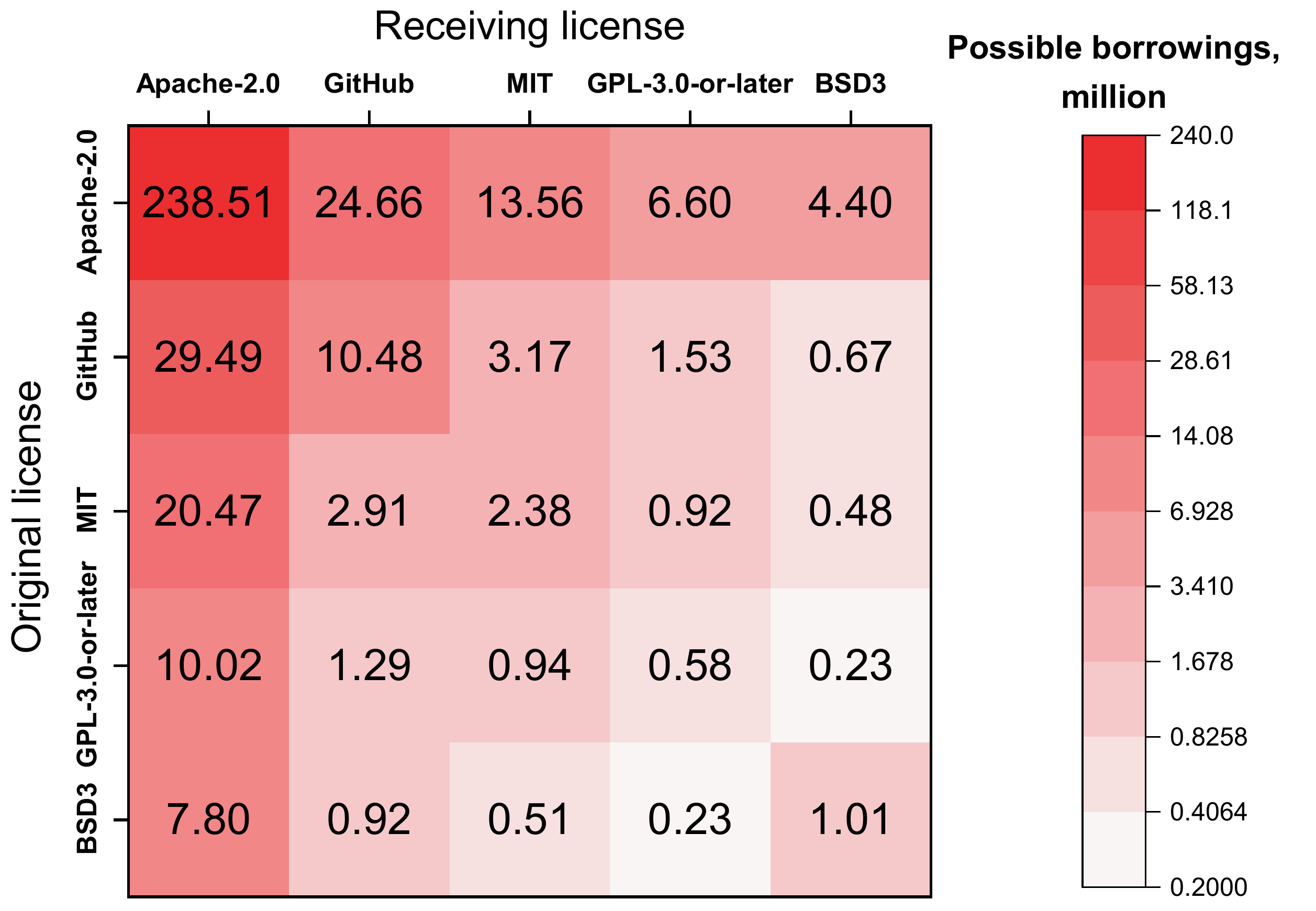}
  \caption{Heatmap of possible borrowings between most frequent licenses with logarithmic coloring, in millions.}
  \label{heatmap}
\end{figure}

Ten most frequent license pairs are presented in Figure~\ref{top10borrowings}. Together, these ten pairs of licenses constitute almost 80\% of all possible borrowings, and they all feature prominent and well-known licenses. As can be expected, copying from Apache-2.0 to Apache-2.0 makes up a little more than half of all possible borrowings due to the popularity of the license in the corpus. 

For the five most frequent licenses, we have drawn up a heatmap of possible borrowings between them, presented in Figure~\ref{heatmap}. An interesting feature of this map is its relative symmetry: for every pair of licenses, the number of possible borrowings from A to B and possible borrowings from B to A is at least similar, there are no significant differences. That might indicate that the amount of possible borrowings between licenses is generally dependant only on the popularity of this license and that if the code is being copied between projects, developers do not pay much attention to the licensing. The most asymmetric license is, once again, Apache-2.0, the possible borrowings to which somewhat outweigh the possible borrowings from it.

The rest of the possible license pairs (that together make up the remaining 1\% of the possible borrowings) were labeled as follows: if either the original file or the receiving file are unlicensed, the pair is deemed prohibited, in all other cases it is considered permitting. Overall, out of 5,471 possible license pairs 5,146 were considered permitting and 325 were considered prohibiting. Please note that in this case, the original or receiving entry could have been a double-license, etc.: for example, if a file with a double license borrows from a file with another double license, it does not make much sense to count it as all four possible combinations.

\begin{figure}
  \centering
  \includegraphics[width=2.8in]{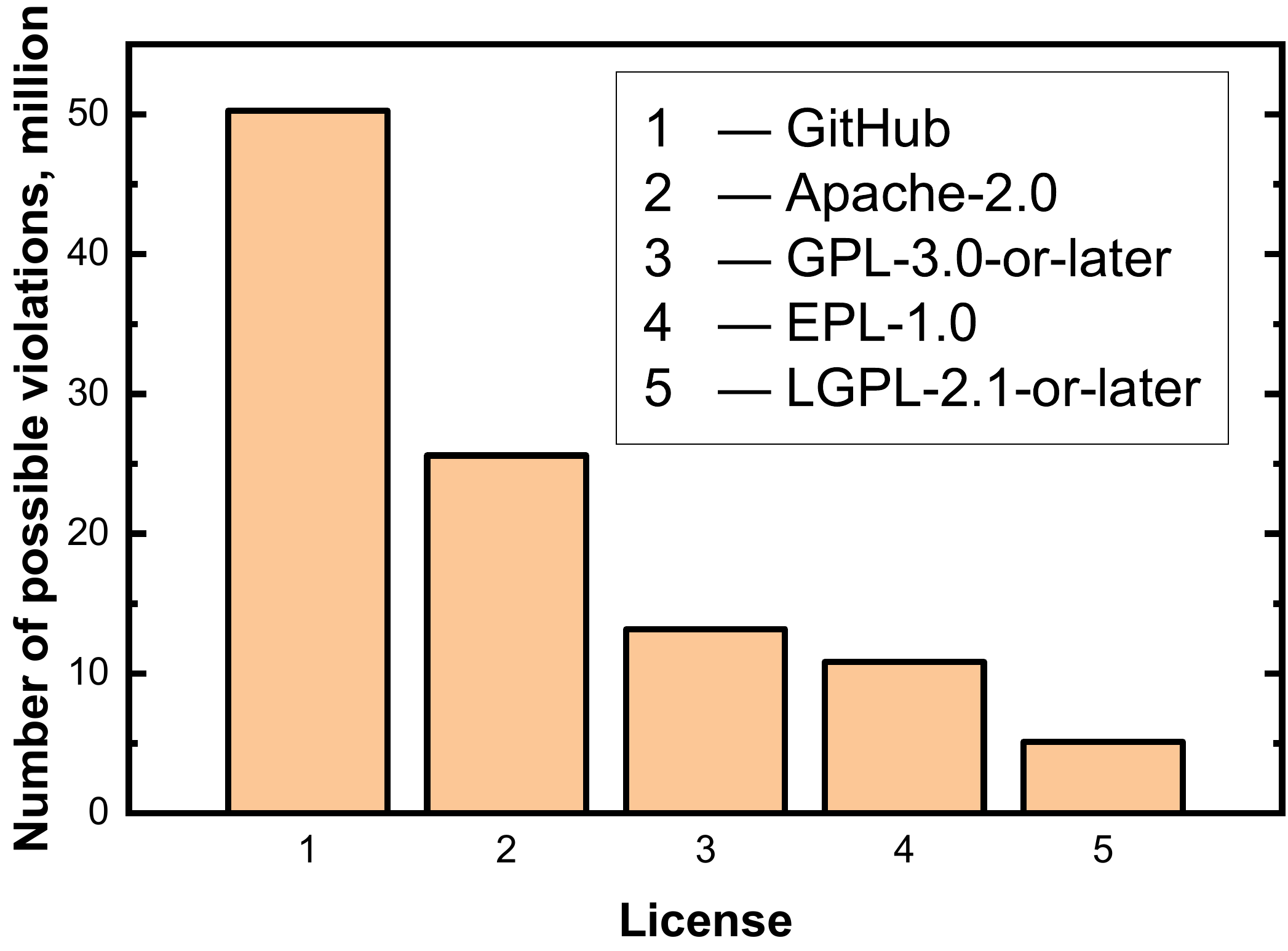}
  \caption{Top five licenses with the most amount of violated blocks of code, i.e. licenses, which act as the source of the largest amount of possible violations.}
  \label{top5violated}
\end{figure}

\begin{figure}
  \centering
  \includegraphics[width=2.8in]{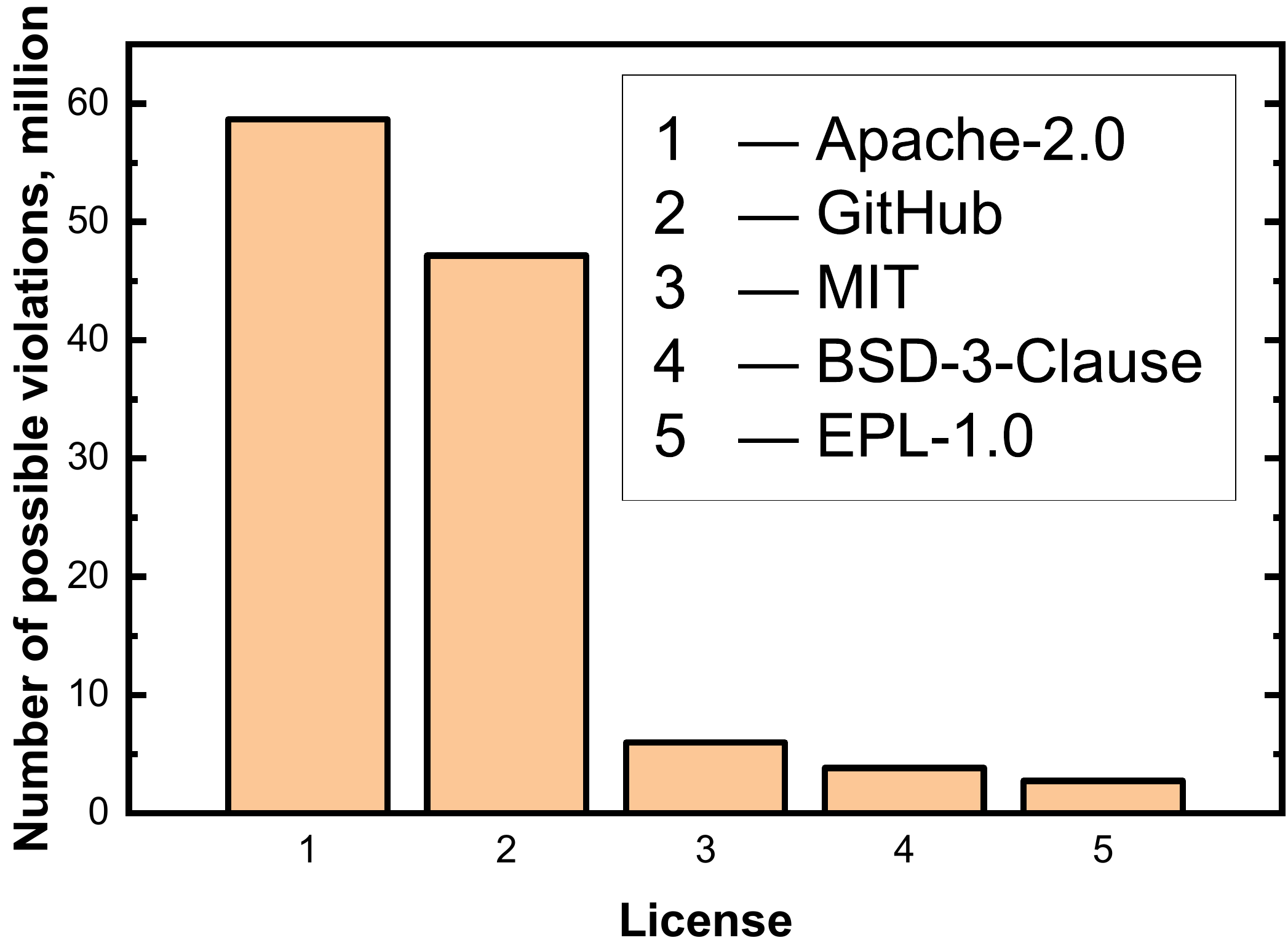}
  \caption{Top five licenses with the most amount of violating pieces of code, i.e. licenses, which act as the receivers of the largest amount of possible violations.}
  \label{top5violating}
\end{figure}

\subsection{Possible license violations}

In this section we analyze possible borrowings that violate the original license. Overall, after the labeling was complete, in total, 72.8\% of the borrowings represented legal borrowings and 27.2\% represented possible license violation.

Figure~\ref{top5violated} shows the most violated licenses, that is, licenses that cover the most blocks that are possibly copied with violating this license. It can be seen that still the majority of possible borrowings violate the rights of developers that did not license their code. Among the most violated licenses there is one permissive license --- Apache-2.0 --- but almost all of this is it being copied to files with no license at all, so its presence in this list is a direct result of its popularity.

Figure~\ref{top5violating} shows the most violating licenses, that is, licenses that cover blocks that are possibly copied with violation of the original licenses. Here, Apache-2.0 and GitHub together make 84\% of all receiving licenses due to their popularity.

In both of these cases, it can be noted that the presence of unlicensed code on GitHub poses a large problem. It is not only prevalent, it appears very often in possible borrowings, and developers should pay closer attention to code with no license. The wide use of Apache-2.0 also leads to the conclusion that developers who employ this license in their projects and include others' work, must be more vigilant to the fact that they cannot use code from projects with strong copyleft licenses.

We have also compiled information about \textit{projects} that have the most possible borrowings that violate their licenses. Of the top three, two do not have any license and one is licensed under EPL-1.0.

\subsection{Possible origins of blocks}

\begin{figure}
  \centering
  \includegraphics[width=2.8in]{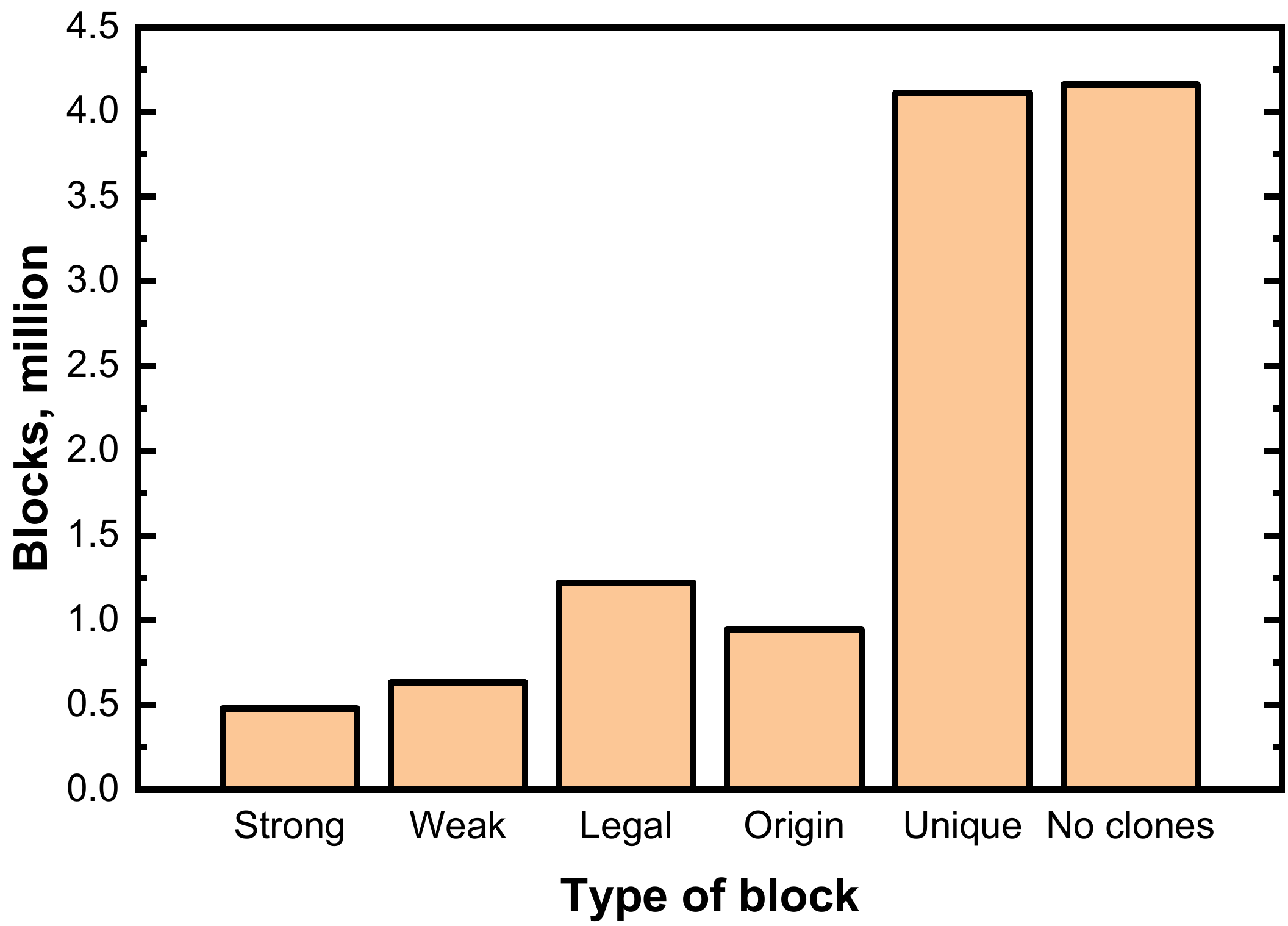}
  \caption{Blocks of code by type in relation to possible license violations.}
  \label{violationfull}
\end{figure}

After that, we have studied the possibility of each block that passes a token length threshold to constitute a violation by running the algorithm described in Section~\ref{sec:analysis_blocks} and divided them into types shown in Figure~\ref{types}. The results are presented in Figure~\ref{violationfull}. 35.4\% have \textit{no clones} at all, meaning that they were tokenized and pass the baseline of length but were not present in SourcererCC's output. 35\% of blocks are \textit{unique}, meaning that they have either clones within their project, or clones between forks, or clones within the same author or organization, but no clones that could constitute a code borrowing or license violation. The remaining 29.6\% of blocks appear in pairs from unrelated projects one way or another.

\textit{Origin} blocks make up 8\% of the blocks, these are the blocks that have clones in other projects, but all of these clones were modified more recently, meaning that this piece of code can only act as an origin point of a possible borrowing. Note that another 1.8\% of blocks also make up a borderline case, where they have valid clones but they are all (usually one or two) have the same day of the last modification. We have evaluated some of such cases and discovered that sometimes such blocks come from projects with the same top contributors.

\begin{figure}
  \centering
  \includegraphics[width=2.8in]{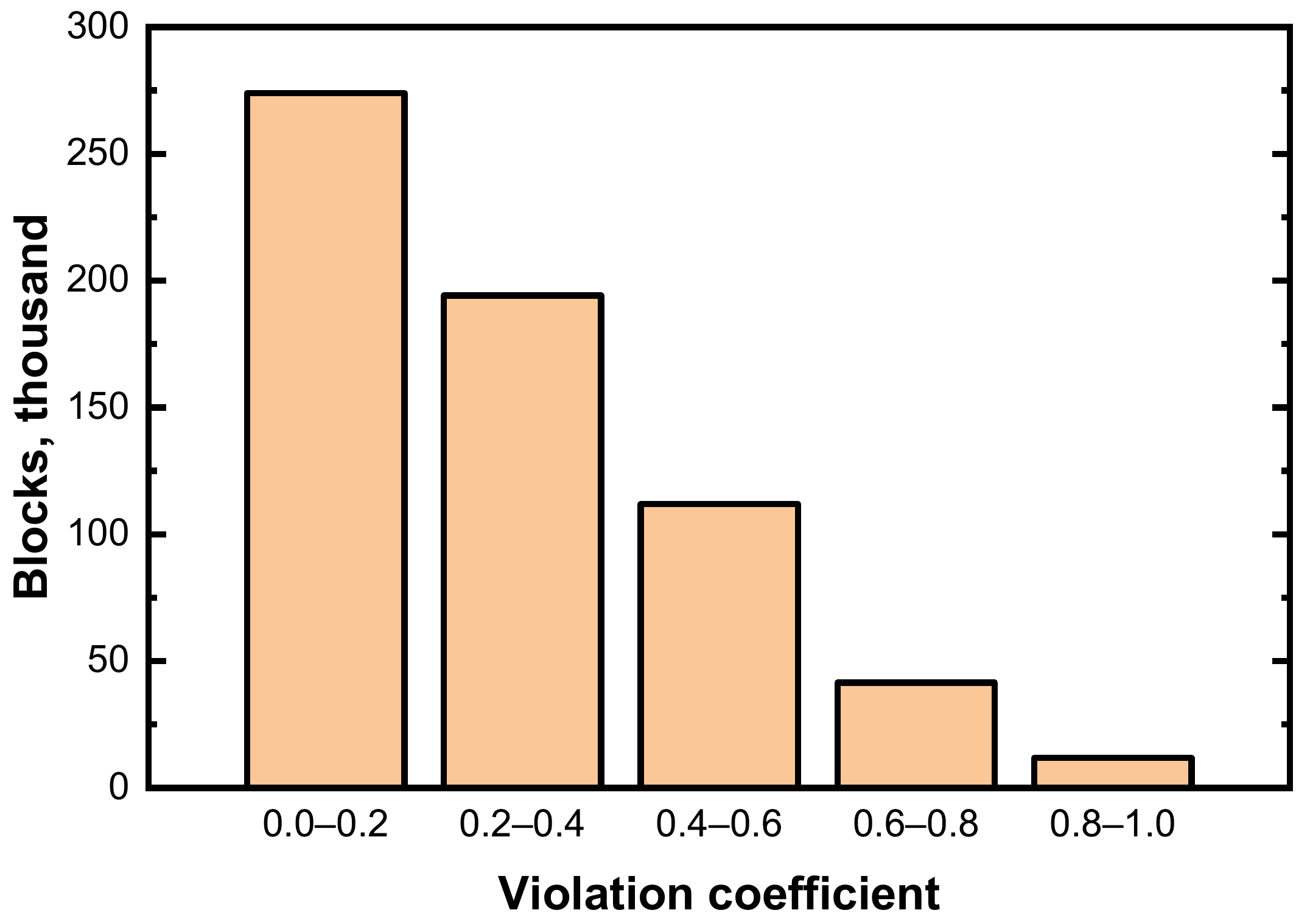}
  \caption{Blocks of code that constitute weak violation by Violation coefficient.}
  \label{violationweak}
\end{figure}

Next, 10.4\% of blocks constitute \textit{legal borrowings}, meaning that they have older clones, but all of their licenses allow this transition. 

Finally, the remaining 9.4\% of blocks constitute possible license violations. Of them, 5.4\% constitute \textit{weak violation}, meaning that only some of their older clones prohibit the possible borrowing, and 4\% constitute \textit{strong violation}, meaning that they have older clones and all of them come from a restricting license. 

Weak violation also differs by the Violation coefficient, as described in Eq.~\ref{eq:1}. Since in general there are more non-violating blocks than violating blocks and the majority of license pairs permit the borrowing, it would make sense for there to be more blocks that constitute weak violation with lower Violation coefficient values. That is exactly the case, as can be seen in Figure~\ref{violationweak}.

Overall, our study has shown that despite a large percentage of unique fragments of code, license violations are still a significant problem in modern software engineering. Over a million blocks of code in our dataset might constitute violations with various degrees of possibility.

\section{Threats to Validity}\label{threats}
The general approach and a broad scope of this study leads to certain threats to validity.

Internal threats to validity are:

\begin{itemize}
    \item There is no consensus in the software engineering community as to what exactly a code clone is and what its boundaries are. Specifically for token-based clone detection, there is no agreement about the minimum token threshold. We have chosen 19 tokens based on our tests, which complies with recent research~\cite{saini2018cloned}.
    \item One of the goals of this study is to examine possible borrowings license-wise, but we cannot make a case whether any of the detected clones were actually plagiarized. However, we believe that the nature of the proposed approach can give a good idea of the general picture of possible borrowings in code and the correlations between the most popular licenses.
    \item In our analysis, we assume that if a file has a license within itself, it substitutes the project's license. The reason for that is that in a majority of cases when they coincide, it does not matter, and when they differ, it is for a reason and it is better to assume that the file's license might bring restriction on its use. However, it is possible to have a more complex relationship between the licenses of the file and the project, including dual-licensing.
    \item While the algorithm for calculating the block's last time of modification using modes allows to estimate borrowings more realistically, some possibilities are not accounted for. For example, the owner of the repository might have borrowed the code from the project with a permitting license, but afterwards that project might have been deleted or made private, and therefore was not collected for our dataset, or the project might have changed its license to a stricter one. Finally, the code might have been copied from outside the PGA altogether.
\end{itemize}

External threats to validity are:
\begin{itemize}
    \item There are other platforms for open-source software, in our work we only concentrate on GitHub, since it is the most popular one.
    \item The dataset was collected on the base of the Public Git Archive, which only includes projects with at least 50 stars. We believe that this is an acceptable restriction, because it makes the operation time of clone detection feasible and also because license violations in larger and more popular projects have more consequences.
\end{itemize}

Construct threats to validity:
\begin{itemize}
    \item SourcererCC detects some false positive clones, however, we believe that the precision of 90\% is suitable for such a large-scale study that demands good scalability.
    \item Both of the tools that are implemented for license detection are not perfect and can detect them wrongfully. However, these are the state-of-the-art tools and they are good at detecting the most popular licenses (Apache, GPL, etc.), which comprise the majority of licensed files in our corpus, moreover, manual checks were applied to the detected licenses to ensure the correct results.
\end{itemize}

All of these threats are important to note, but we believe they do not invalidate the main findings in the research, namely, the scope and the general picture of license distribution, possible borrowings, and possible licence violations in Java code on a large open-source software platform.

\section{Conclusion and Future work}\label{conclusion}

In this paper,  we conduct a study of possible block-level code borrowing and license violations in the Java corpus of GitHub. In our research, we have discovered that 48.2\% of the 1,163,989,420 detected clone pairs are inter-project (and are therefore relevant to our analysis).

From the standpoint of licensing, our study has detected 94 different unique licenses in the dataset. They are distributed very unequally, with the most frequent ones being the most famous ones like permissive Apache-2.0, MIT, and BSD-3-Clause, as well as more restrictive ones like GPL licenses, LGPL licenses, and AGPL license. A very important specific case has to do with the absence of a license: at least 9.1\% of the checked files are covered by no license at all, which leads to a lot of possible violations. The amount of files with several licenses is much lower.

The analysis of possible borrowings has shown that more than half of them falls within Apache-2.0, which is also the most popular license we have discovered by far. One of our findings is that unlicensed code is the most probable to be violated, we have drawn up a heatmap of possible borrowings between the most famous licenses.

Finally, we have studied specific blocks to determine their possibility of constituting licensing violations. We have discovered that 35.4\% of blocks have no clones at all and that 9.4\% of blocks might constitute possible violations, of which 4\% --- strongly, meaning that all of their older clones come from restrictive licenses.

This research can be expanded in a variety of different directions, here are some of them:
\begin{itemize}
    \item It is possible to broaden the dataset to include less popular code by adding the repositories with less than 50 stars. It might be of interest to check whether the fraction of potential violations in such a dataset is higher or not. 
    \item One might also consider other open-source software platforms and other programming languages, where the distribution of licenses might be different.
    \item Another interesting relevant research might be conducted on the relations between the file's license and the project's license. In our research, we have assumed the file's license (if there is any) over the project's license, but we did not study the cases where the file's license and the project's license are not compatible in the first place, since it does not relate to possible borrowing. One can search for such cases and study their prevalence.
\end{itemize}

Overall, with the omnipresent development of open-source software and the rapid expansion of software engineering, the issue of code borrowing and license violations will only become more relevant. Conducting studies on the current state of affairs is an integral part in keeping track of this problem, and we hope that developers will be more attentive towards the licensing of open-source software.

\bibliographystyle{ACM-Reference-Format}
\bibliography{cites} 


\begin{thebibliography}{33}


\ifx \showCODEN    \undefined \def \showCODEN     #1{\unskip}     \fi
\ifx \showDOI      \undefined \def \showDOI       #1{#1}\fi
\ifx \showISBNx    \undefined \def \showISBNx     #1{\unskip}     \fi
\ifx \showISBNxiii \undefined \def \showISBNxiii  #1{\unskip}     \fi
\ifx \showISSN     \undefined \def \showISSN      #1{\unskip}     \fi
\ifx \showLCCN     \undefined \def \showLCCN      #1{\unskip}     \fi
\ifx \shownote     \undefined \def \shownote      #1{#1}          \fi
\ifx \showarticletitle \undefined \def \showarticletitle #1{#1}   \fi
\ifx \showURL      \undefined \def \showURL       {\relax}        \fi
\providecommand\bibfield[2]{#2}
\providecommand\bibinfo[2]{#2}
\providecommand\natexlab[1]{#1}
\providecommand\showeprint[2][]{arXiv:#2}

\bibitem[\protect\citeauthoryear{Ain, Butt, Anwar, Azam, and Maqbool}{Ain
  et~al\mbox{.}}{2019}]%
        {ain2019recent}
\bibfield{author}{\bibinfo{person}{Qurat~Ul Ain}, \bibinfo{person}{Wasi~Haider
  Butt}, \bibinfo{person}{Muhamad~Waseem Anwar}, \bibinfo{person}{Farooque
  Azam}, {and} \bibinfo{person}{Bilal Maqbool}.}
  \bibinfo{year}{2019}\natexlab{}.
\newblock \showarticletitle{Recent Advancements in Code Clone
  Detection--Techniques and Tools}.
\newblock \bibinfo{journal}{\emph{IEEE Access}}  \bibinfo{volume}{7}
  (\bibinfo{year}{2019}).
\newblock


\bibitem[\protect\citeauthoryear{Almeida, Murphy, Wilson, and Hoye}{Almeida
  et~al\mbox{.}}{2017}]%
        {almeida2017software}
\bibfield{author}{\bibinfo{person}{Daniel~A Almeida}, \bibinfo{person}{Gail~C
  Murphy}, \bibinfo{person}{Greg Wilson}, {and} \bibinfo{person}{Mike Hoye}.}
  \bibinfo{year}{2017}\natexlab{}.
\newblock \showarticletitle{Do software developers understand open source
  licenses?}. In \bibinfo{booktitle}{\emph{Proceedings of the 25th
  International Conference on Program Comprehension}}. \bibinfo{pages}{1--11}.
\newblock


\bibitem[\protect\citeauthoryear{Baker}{Baker}{1995}]%
        {baker1995finding}
\bibfield{author}{\bibinfo{person}{Brenda~S Baker}.}
  \bibinfo{year}{1995}\natexlab{}.
\newblock \showarticletitle{On finding duplication and near-duplication in
  large software systems}. In \bibinfo{booktitle}{\emph{Proceedings of 2nd
  Working Conference on Reverse Engineering}}. \bibinfo{pages}{86--95}.
\newblock


\bibitem[\protect\citeauthoryear{Bruntink, Van~Deursen, Van~Engelen, and
  Tourwe}{Bruntink et~al\mbox{.}}{2005}]%
        {bruntink2005use}
\bibfield{author}{\bibinfo{person}{Magiel Bruntink}, \bibinfo{person}{Arie
  Van~Deursen}, \bibinfo{person}{Remco Van~Engelen}, {and} \bibinfo{person}{Tom
  Tourwe}.} \bibinfo{year}{2005}\natexlab{}.
\newblock \showarticletitle{On the use of clone detection for identifying
  crosscutting concern code}.
\newblock \bibinfo{journal}{\emph{IEEE Transactions on Software Engineering}}
  \bibinfo{volume}{31}, \bibinfo{number}{10} (\bibinfo{year}{2005}),
  \bibinfo{pages}{804--818}.
\newblock


\bibitem[\protect\citeauthoryear{Chatterji, Carver, Kraft, and
  Harder}{Chatterji et~al\mbox{.}}{2013}]%
        {chatterji2013effects}
\bibfield{author}{\bibinfo{person}{Debarshi Chatterji},
  \bibinfo{person}{Jeffrey~C Carver}, \bibinfo{person}{Nicholas~A Kraft}, {and}
  \bibinfo{person}{Jan Harder}.} \bibinfo{year}{2013}\natexlab{}.
\newblock \showarticletitle{Effects of cloned code on software maintainability:
  A replicated developer study}. In \bibinfo{booktitle}{\emph{2013 20th Working
  Conference on Reverse Engineering (WCRE)}}. \bibinfo{pages}{112--121}.
\newblock


\bibitem[\protect\citeauthoryear{Chatterji, Carver, Massengil, Oslin, and
  Kraft}{Chatterji et~al\mbox{.}}{2011}]%
        {chatterji2011measuring}
\bibfield{author}{\bibinfo{person}{Debarshi Chatterji},
  \bibinfo{person}{Jeffrey~C Carver}, \bibinfo{person}{Beverly Massengil},
  \bibinfo{person}{Jason Oslin}, {and} \bibinfo{person}{Nicholas~A Kraft}.}
  \bibinfo{year}{2011}\natexlab{}.
\newblock \showarticletitle{Measuring the efficacy of code clone information in
  a bug localization task: An empirical study}. In
  \bibinfo{booktitle}{\emph{2011 International Symposium on Empirical Software
  Engineering and Measurement}}. \bibinfo{pages}{20--29}.
\newblock


\bibitem[\protect\citeauthoryear{German, Di~Penta, and Davies}{German
  et~al\mbox{.}}{2010a}]%
        {german2010understanding}
\bibfield{author}{\bibinfo{person}{Daniel~M German},
  \bibinfo{person}{Massimiliano Di~Penta}, {and} \bibinfo{person}{Julius
  Davies}.} \bibinfo{year}{2010}\natexlab{a}.
\newblock \showarticletitle{Understanding and auditing the licensing of open
  source software distributions}. In \bibinfo{booktitle}{\emph{2010 IEEE 18th
  International Conference on Program Comprehension}}. \bibinfo{pages}{84--93}.
\newblock


\bibitem[\protect\citeauthoryear{German, Di~Penta, Gueheneuc, and
  Antoniol}{German et~al\mbox{.}}{2009}]%
        {german2009code}
\bibfield{author}{\bibinfo{person}{Daniel~M German},
  \bibinfo{person}{Massimiliano Di~Penta}, \bibinfo{person}{Yann-Gael
  Gueheneuc}, {and} \bibinfo{person}{Giuliano Antoniol}.}
  \bibinfo{year}{2009}\natexlab{}.
\newblock \showarticletitle{Code siblings: Technical and legal implications of
  copying code between applications}. In \bibinfo{booktitle}{\emph{2009 6th
  IEEE International Working Conference on Mining Software Repositories}}.
  \bibinfo{pages}{81--90}.
\newblock


\bibitem[\protect\citeauthoryear{German, Manabe, and Inoue}{German
  et~al\mbox{.}}{2010b}]%
        {Ninka}
\bibfield{author}{\bibinfo{person}{Daniel~M. German}, \bibinfo{person}{Yuki
  Manabe}, {and} \bibinfo{person}{Katsuro Inoue}.}
  \bibinfo{year}{2010}\natexlab{b}.
\newblock \showarticletitle{A Sentence-matching Method for Automatic License
  Identification of Source Code Files}. In
  \bibinfo{booktitle}{\emph{Proceedings of the IEEE/ACM International
  Conference on Automated Software Engineering}} \emph{(\bibinfo{series}{ASE
  '10})}. \bibinfo{pages}{437--446}.
\newblock


\bibitem[\protect\citeauthoryear{Gousios}{Gousios}{2013}]%
        {GHTorrent}
\bibfield{author}{\bibinfo{person}{Georgios Gousios}.}
  \bibinfo{year}{2013}\natexlab{}.
\newblock \showarticletitle{The GHTorrent dataset and tool suite}. In
  \bibinfo{booktitle}{\emph{Proceedings of the 10th Working Conference on
  Mining Software Repositories}} \emph{(\bibinfo{series}{MSR '13})}.
  \bibinfo{pages}{233--236}.
\newblock


\bibitem[\protect\citeauthoryear{Krinke, Gold, Jia, and Binkley}{Krinke
  et~al\mbox{.}}{2010a}]%
        {krinke2010cloning}
\bibfield{author}{\bibinfo{person}{Jens Krinke}, \bibinfo{person}{Nicolas
  Gold}, \bibinfo{person}{Yue Jia}, {and} \bibinfo{person}{David Binkley}.}
  \bibinfo{year}{2010}\natexlab{a}.
\newblock \showarticletitle{Cloning and copying between GNOME projects}. In
  \bibinfo{booktitle}{\emph{2010 7th IEEE Working Conference on Mining Software
  Repositories (MSR 2010)}}. \bibinfo{pages}{98--101}.
\newblock


\bibitem[\protect\citeauthoryear{Krinke, Gold, Jia, and Binkley}{Krinke
  et~al\mbox{.}}{2010b}]%
        {krinke2010distinguishing}
\bibfield{author}{\bibinfo{person}{Jens Krinke}, \bibinfo{person}{Nicolas
  Gold}, \bibinfo{person}{Yue Jia}, {and} \bibinfo{person}{David Binkley}.}
  \bibinfo{year}{2010}\natexlab{b}.
\newblock \showarticletitle{Distinguishing copies from originals in software
  clones}. In \bibinfo{booktitle}{\emph{Proceedings of the 4th International
  Workshop on Software Clones}}. \bibinfo{pages}{41--48}.
\newblock


\bibitem[\protect\citeauthoryear{Lopes, Maj, Martins, Saini, Yang, Zitny,
  Sajnani, and Vitek}{Lopes et~al\mbox{.}}{2017}]%
        {lopes2017dejavu}
\bibfield{author}{\bibinfo{person}{Cristina~V Lopes}, \bibinfo{person}{Petr
  Maj}, \bibinfo{person}{Pedro Martins}, \bibinfo{person}{Vaibhav Saini},
  \bibinfo{person}{Di Yang}, \bibinfo{person}{Jakub Zitny},
  \bibinfo{person}{Hitesh Sajnani}, {and} \bibinfo{person}{Jan Vitek}.}
  \bibinfo{year}{2017}\natexlab{}.
\newblock \showarticletitle{D{\'e}j{\`a}Vu: a map of code duplicates on
  GitHub}.
\newblock \bibinfo{journal}{\emph{Proceedings of the ACM on Programming
  Languages}} \bibinfo{volume}{1}, \bibinfo{number}{OOPSLA}
  (\bibinfo{year}{2017}), \bibinfo{pages}{84}.
\newblock


\bibitem[\protect\citeauthoryear{Markovtsev and Long}{Markovtsev and
  Long}{2018}]%
        {PGA}
\bibfield{author}{\bibinfo{person}{Vadim Markovtsev} {and}
  \bibinfo{person}{Waren Long}.} \bibinfo{year}{2018}\natexlab{}.
\newblock \showarticletitle{Public Git Archive: a Big Code dataset for all}.
\newblock \bibinfo{journal}{\emph{CoRR}}  \bibinfo{volume}{abs/1803.10144}
  (\bibinfo{year}{2018}).
\newblock


\bibitem[\protect\citeauthoryear{Mathur, Choudhary, Vashist, Thies, and
  Thilagam}{Mathur et~al\mbox{.}}{2012}]%
        {mathur2012empirical}
\bibfield{author}{\bibinfo{person}{Arunesh Mathur}, \bibinfo{person}{Harshal
  Choudhary}, \bibinfo{person}{Priyank Vashist}, \bibinfo{person}{William
  Thies}, {and} \bibinfo{person}{Santhi Thilagam}.}
  \bibinfo{year}{2012}\natexlab{}.
\newblock \showarticletitle{An empirical study of license violations in open
  source projects}. In \bibinfo{booktitle}{\emph{2012 35th Annual IEEE Software
  Engineering Workshop}}. \bibinfo{pages}{168--176}.
\newblock


\bibitem[\protect\citeauthoryear{Mockus}{Mockus}{2007}]%
        {mockus2007large}
\bibfield{author}{\bibinfo{person}{Audris Mockus}.}
  \bibinfo{year}{2007}\natexlab{}.
\newblock \showarticletitle{Large-scale code reuse in open source software}. In
  \bibinfo{booktitle}{\emph{First International Workshop on Emerging Trends in
  FLOSS Research and Development (FLOSS'07: ICSE Workshops 2007)}}.
  \bibinfo{pages}{7--7}.
\newblock


\bibitem[\protect\citeauthoryear{Mockus}{Mockus}{2009}]%
        {mockus2009amassing}
\bibfield{author}{\bibinfo{person}{Audris Mockus}.}
  \bibinfo{year}{2009}\natexlab{}.
\newblock \showarticletitle{Amassing and indexing a large sample of version
  control systems: Towards the census of public source code history}. In
  \bibinfo{booktitle}{\emph{2009 6th IEEE International Working Conference on
  Mining Software Repositories}}. \bibinfo{pages}{11--20}.
\newblock


\bibitem[\protect\citeauthoryear{Mondal, Roy, Roy, and Schneider}{Mondal
  et~al\mbox{.}}{2019}]%
        {mondal2019investigating}
\bibfield{author}{\bibinfo{person}{Manishankar Mondal}, \bibinfo{person}{Banani
  Roy}, \bibinfo{person}{Chanchal~K Roy}, {and} \bibinfo{person}{Kevin~A
  Schneider}.} \bibinfo{year}{2019}\natexlab{}.
\newblock \showarticletitle{Investigating Context Adaptation Bugs in Code
  Clones}. In \bibinfo{booktitle}{\emph{2019 IEEE International Conference on
  Software Maintenance and Evolution (ICSME)}}. \bibinfo{pages}{157--168}.
\newblock


\bibitem[\protect\citeauthoryear{Ossher, Sajnani, and Lopes}{Ossher
  et~al\mbox{.}}{2011}]%
        {ossher2011file}
\bibfield{author}{\bibinfo{person}{Joel Ossher}, \bibinfo{person}{Hitesh
  Sajnani}, {and} \bibinfo{person}{Cristina Lopes}.}
  \bibinfo{year}{2011}\natexlab{}.
\newblock \showarticletitle{File cloning in open source java projects: The
  good, the bad, and the ugly}. In \bibinfo{booktitle}{\emph{2011 27th IEEE
  International Conference on Software Maintenance (ICSM)}}.
  \bibinfo{pages}{283--292}.
\newblock


\bibitem[\protect\citeauthoryear{Rattan, Bhatia, and Singh}{Rattan
  et~al\mbox{.}}{2013}]%
        {rattan2013software}
\bibfield{author}{\bibinfo{person}{Dhavleesh Rattan}, \bibinfo{person}{Rajesh
  Bhatia}, {and} \bibinfo{person}{Maninder Singh}.}
  \bibinfo{year}{2013}\natexlab{}.
\newblock \showarticletitle{Software clone detection: A systematic review}.
\newblock \bibinfo{journal}{\emph{Information and Software Technology}}
  \bibinfo{volume}{55}, \bibinfo{number}{7} (\bibinfo{year}{2013}),
  \bibinfo{pages}{1165--1199}.
\newblock


\bibitem[\protect\citeauthoryear{Romansky, Chen, Malhotra, and Hindle}{Romansky
  et~al\mbox{.}}{2018}]%
        {romansky2018sourcerer}
\bibfield{author}{\bibinfo{person}{Stephen Romansky}, \bibinfo{person}{Cheng
  Chen}, \bibinfo{person}{Baljeet Malhotra}, {and} \bibinfo{person}{Abram
  Hindle}.} \bibinfo{year}{2018}\natexlab{}.
\newblock \showarticletitle{Sourcerer's Apprentice and the study of code
  snippet migration}.
\newblock \bibinfo{journal}{\emph{arXiv preprint arXiv:1808.00106}}
  (\bibinfo{year}{2018}).
\newblock


\bibitem[\protect\citeauthoryear{Roy and Cordy}{Roy and Cordy}{2008a}]%
        {roy2008empirical}
\bibfield{author}{\bibinfo{person}{Chanchal~K Roy} {and}
  \bibinfo{person}{James~R Cordy}.} \bibinfo{year}{2008}\natexlab{a}.
\newblock \showarticletitle{An empirical study of function clones in open
  source software}. In \bibinfo{booktitle}{\emph{2008 15th Working Conference
  on Reverse Engineering}}. \bibinfo{pages}{81--90}.
\newblock


\bibitem[\protect\citeauthoryear{Roy and Cordy}{Roy and Cordy}{2008b}]%
        {roy2008nicad}
\bibfield{author}{\bibinfo{person}{Chanchal~K Roy} {and}
  \bibinfo{person}{James~R Cordy}.} \bibinfo{year}{2008}\natexlab{b}.
\newblock \showarticletitle{NICAD: Accurate detection of near-miss intentional
  clones using flexible pretty-printing and code normalization}. In
  \bibinfo{booktitle}{\emph{2008 16th iEEE international conference on program
  comprehension}}. \bibinfo{pages}{172--181}.
\newblock


\bibitem[\protect\citeauthoryear{Roy and Cordy}{Roy and Cordy}{2010}]%
        {roy2010near}
\bibfield{author}{\bibinfo{person}{Chanchal~K Roy} {and}
  \bibinfo{person}{James~R Cordy}.} \bibinfo{year}{2010}\natexlab{}.
\newblock \showarticletitle{Near-miss function clones in open source software:
  an empirical study}.
\newblock \bibinfo{journal}{\emph{Journal of Software Maintenance and
  Evolution: Research and Practice}} \bibinfo{volume}{22}, \bibinfo{number}{3}
  (\bibinfo{year}{2010}), \bibinfo{pages}{165--189}.
\newblock


\bibitem[\protect\citeauthoryear{Saini, Sajnani, and Lopes}{Saini
  et~al\mbox{.}}{2018}]%
        {saini2018cloned}
\bibfield{author}{\bibinfo{person}{Vaibhav Saini}, \bibinfo{person}{Hitesh
  Sajnani}, {and} \bibinfo{person}{Cristina Lopes}.}
  \bibinfo{year}{2018}\natexlab{}.
\newblock \showarticletitle{Cloned and non-cloned Java methods: a comparative
  study}.
\newblock \bibinfo{journal}{\emph{Empirical Software Engineering}}
  \bibinfo{volume}{23}, \bibinfo{number}{4} (\bibinfo{year}{2018}),
  \bibinfo{pages}{2232--2278}.
\newblock


\bibitem[\protect\citeauthoryear{Sajnani, Saini, Svajlenko, Roy, and
  Lopes}{Sajnani et~al\mbox{.}}{2015}]%
        {SourcererCC}
\bibfield{author}{\bibinfo{person}{Hitesh Sajnani}, \bibinfo{person}{Vaibhav
  Saini}, \bibinfo{person}{Jeffrey Svajlenko}, \bibinfo{person}{Chanchal~K.
  Roy}, {and} \bibinfo{person}{Cristina~V. Lopes}.}
  \bibinfo{year}{2015}\natexlab{}.
\newblock \showarticletitle{SourcererCC: Scaling Code Clone Detection to Big
  Code}.
\newblock \bibinfo{journal}{\emph{CoRR}}  \bibinfo{volume}{abs/1512.06448}
  (\bibinfo{year}{2015}).
\newblock


\bibitem[\protect\citeauthoryear{Vendome}{Vendome}{2015}]%
        {vendome2015large}
\bibfield{author}{\bibinfo{person}{Christopher Vendome}.}
  \bibinfo{year}{2015}\natexlab{}.
\newblock \showarticletitle{A large scale study of license usage on GitHub}. In
  \bibinfo{booktitle}{\emph{Proceedings of the 37th International Conference on
  Software Engineering-Volume 2}}. \bibinfo{pages}{772--774}.
\newblock


\bibitem[\protect\citeauthoryear{Vendome, Linares-Vasquez, Bavota, Di~Penta,
  German, and Poshyvanyk}{Vendome et~al\mbox{.}}{2015}]%
        {vendome2015and}
\bibfield{author}{\bibinfo{person}{Christopher Vendome}, \bibinfo{person}{Mario
  Linares-Vasquez}, \bibinfo{person}{Gabriele Bavota},
  \bibinfo{person}{Massimiliano Di~Penta}, \bibinfo{person}{Daniel~M German},
  {and} \bibinfo{person}{Denys Poshyvanyk}.} \bibinfo{year}{2015}\natexlab{}.
\newblock \showarticletitle{When and why developers adopt and change software
  licenses}. In \bibinfo{booktitle}{\emph{2015 IEEE International Conference on
  Software Maintenance and Evolution (ICSME)}}. \bibinfo{pages}{31--40}.
\newblock


\bibitem[\protect\citeauthoryear{von Krogh, Spaeth, and Haefliger}{von Krogh
  et~al\mbox{.}}{2005}]%
        {haefliger2006knowledge}
\bibfield{author}{\bibinfo{person}{Georg von Krogh}, \bibinfo{person}{Sebastian
  Spaeth}, {and} \bibinfo{person}{Stefan Haefliger}.}
  \bibinfo{year}{2005}\natexlab{}.
\newblock \showarticletitle{Knowledge reuse in open source software: An
  exploratory study of 15 open source projects}. In
  \bibinfo{booktitle}{\emph{Proceedings of the 38th Annual Hawaii International
  Conference on System Sciences}}. \bibinfo{pages}{198b--198b}.
\newblock


\bibitem[\protect\citeauthoryear{Wu, Manabe, Kanda, German, and Inoue}{Wu
  et~al\mbox{.}}{2015}]%
        {wu2015method}
\bibfield{author}{\bibinfo{person}{Yuhao Wu}, \bibinfo{person}{Yuki Manabe},
  \bibinfo{person}{Tetsuya Kanda}, \bibinfo{person}{Daniel~M German}, {and}
  \bibinfo{person}{Katsuro Inoue}.} \bibinfo{year}{2015}\natexlab{}.
\newblock \showarticletitle{A method to detect license inconsistencies in
  large-scale open source projects}. In \bibinfo{booktitle}{\emph{2015 IEEE/ACM
  12th Working Conference on Mining Software Repositories}}.
  \bibinfo{pages}{324--333}.
\newblock


\bibitem[\protect\citeauthoryear{Wu, Manabe, Kanda, German, and Inoue}{Wu
  et~al\mbox{.}}{2017}]%
        {wu2017analysis}
\bibfield{author}{\bibinfo{person}{Yuhao Wu}, \bibinfo{person}{Yuki Manabe},
  \bibinfo{person}{Tetsuya Kanda}, \bibinfo{person}{Daniel~M German}, {and}
  \bibinfo{person}{Katsuro Inoue}.} \bibinfo{year}{2017}\natexlab{}.
\newblock \showarticletitle{Analysis of license inconsistency in large
  collections of open source projects}.
\newblock \bibinfo{journal}{\emph{Empirical Software Engineering}}
  \bibinfo{volume}{22}, \bibinfo{number}{3} (\bibinfo{year}{2017}),
  \bibinfo{pages}{1194--1222}.
\newblock


\bibitem[\protect\citeauthoryear{Yang, Martins, Saini, and Lopes}{Yang
  et~al\mbox{.}}{2017}]%
        {yang2017stack}
\bibfield{author}{\bibinfo{person}{Di Yang}, \bibinfo{person}{Pedro Martins},
  \bibinfo{person}{Vaibhav Saini}, {and} \bibinfo{person}{Cristina Lopes}.}
  \bibinfo{year}{2017}\natexlab{}.
\newblock \showarticletitle{Stack overflow in GitHub: any snippets there?}. In
  \bibinfo{booktitle}{\emph{2017 IEEE/ACM 14th International Conference on
  Mining Software Repositories (MSR)}}. \bibinfo{pages}{280--290}.
\newblock


\bibitem[\protect\citeauthoryear{Zhang, Shi, and Zhang}{Zhang
  et~al\mbox{.}}{2010}]%
        {zhang2010automatic}
\bibfield{author}{\bibinfo{person}{Hongyu Zhang}, \bibinfo{person}{Bei Shi},
  {and} \bibinfo{person}{Lu Zhang}.} \bibinfo{year}{2010}\natexlab{}.
\newblock \showarticletitle{Automatic checking of license compliance}. In
  \bibinfo{booktitle}{\emph{2010 IEEE International Conference on Software
  Maintenance}}. \bibinfo{pages}{1--3}.
\newblock


\end{thebibliography}
\end{document}